\documentclass[francais]{gretsi}

\usepackage[utf8]{inputenc}

\usepackage[T1]{fontenc}

\usepackage{amsmath, amssymb, amsfonts}

\usepackage{caption}

\begin{document}


\title{Désentrelacement fréquentiel doux pour les codecs audio neuronaux}

\auteurs{
  \auteur{Benoît}{Giniès}{benoit.ginies@telecom-paris.fr}{}
  \auteur{Xiaoyu}{Bie}{xiaoyu.bie@telecom-paris.fr}{}
  \auteur{Olivier}{Fercoq}{olivier.fercoq@telecom-paris.fr}{}
  \auteur{Gaël}{Richard}{gael.richard@telecom.paris.fr}{}
}

\affils{
  \affil{}{LTCI, T\'el\'ecom Paris, Institut polytechnique de Paris, Palaiseau, France\\
  }
}

\resume{Bien que les modèles basés sur les réseaux de neurones aient permis des avancées significatives dans l'extraction de représentations audio, l'interprétabilité des représentations apprises reste un défi majeur. Pour y remédier, des techniques de désentrelacement ont été intégrées dans les codecs audio neuronaux discrets afin d'imposer une structure aux tokens extraits. Cependant, ces approches sont souvent fortement dépendantes de tâches ou d'ensembles de données spécifiques. Dans ce travail, nous proposons un codec audio neuronal désentrelacé qui tire parti de la décomposition spectrale des signaux temporels pour améliorer l'interprétabilité de la représentation. Des évaluations expérimentales démontrent que notre méthode surpasse un modèle de référence en termes de fidélité de reconstruction et de qualité perceptuelle.}

\abstract{While neural-based models have led to significant advancements in audio feature extraction, the interpretability of the learned representations remains a critical challenge. To address this, disentanglement techniques have been integrated into discrete neural audio codecs to impose structure on the extracted tokens. However, these approaches often exhibit strong dependencies on specific datasets or task formulations. In this work, we propose a disentangled neural audio codec that leverages spectral decomposition of time-domain signals to enhance representation interpretability. Experimental evaluations demonstrate that our method surpasses a state-of-the-art baseline in both reconstruction fidelity and perceptual quality.}

\maketitle


\vfill
\begin{minipage}{\columnwidth}
    \footnotesize Ce travail a été financé par l'Union européenne (ERC, HI-Audio, 101052978). Les points de vue et les opinions exprimés sont ceux des auteurs et ne reflètent pas nécessairement ceux de l'Union européenne ou du Conseil européen de la recherche. Ni l'Union européenne ni l'organisme subventionnaire ne peuvent en être tenus pour responsables.
\end{minipage}

\section{Introduction}

En traitement audio, l'extraction d'une représentation efficace du signal est essentielle. L'objectif principal d'une telle extraction est d'encoder les informations les plus pertinentes sous une forme compacte. Traditionnellement, ce compromis entre exhaustivité des données et efficacité de la représentation a conduit au développement de représentations conçues à la main pour capturer les propriétés significatives du signal. Les représentations temps-fréquence y jouent un rôle crucial, avec des approches largement reconnues allant de la transformée de Fourier à court terme (TFCT) à des représentations perceptives exploitant l'échelle Mel par exemple. \cite{stevens1937scale} 

L'utilisation de réseaux neuronaux pour l'extraction de représentations dans le traitement de l'image et de l'audio a connu un essor important avec l'introduction de l'architecture encodeur-décodeur, appliquée dans le modèle VAE \cite{kingma_auto-encoding_2022}, puis étendue avec une étape de quantification dans le modèle VQ-VAE \cite{oord_neural_2018}. En prolongement, le quantificateur vectoriel résiduel (RVQ) utilisé dans \cite{defossez_high_2022, kumar_high-fidelity_2023}, est venu améliorer la reconstruction. Une relaxation de la quantification a été étudié et diverses structures de quantificateurs ont été développées dans \cite{takida_hq-vae_2023}. L'utilisation de représentations discrètes présente un intérêt notable pour la synthèse vocale \cite{wang_neural_2023} ou le transfert de timbre musical~\cite{Cifka2021} par exemple.

Le succès des codecs audio neuronaux a suscité un intérêt croissant pour le désentrelacement des représentations latentes, avec des méthodes souvent adaptées à des applications spécifiques. Par exemple, pour la conversion de la voix chantée, Takahashi \& al. \cite{takahashi_hierarchical_2021} intègre le module VQ avec des codeurs de hauteur et d'amplitude. En synthèse vocale, Adam \& al. \cite{polyak2021speech} renforce le désentrelacement du contenu, de la hauteur et de l'identité du locuteur grâce à des structures d'extraction de représentations spécialisées, tandis que Ju \& al. \cite{ju_naturalspeech_2024} allouent des quantificateurs pour capturer la prosodie, le contexte et les détails acoustiques de la parole, complétés par un encodeur de timbre. L'entraînement de ces quantificateurs est guidé par des tâches de supervision appropriées, idée qui a également été appliquée à la séparation des sources~\cite{bie2024learning} par exemple. Bien que nombre de ces approches soient spécifiques à une application, certaines stratégies plus générales ont été proposées. Par exemple, \cite{hsu_disentanglement_nodate} décrit une méthode visant à encourager le désentrelacement des représentations latentes discrètes, et Luo \& al. \cite{luo_gull_2024} présentent une approche générique avec un codec multibande, où chaque bande de fréquence du spectrogramme d'entrée est traitée séparément.

Dans cet article, nous présentons un nouveau codec neuronal qui génère une représentation discrète et désentrelacée en fréquence. Le modèle de codec proposé fonctionne sur plusieurs fréquences d'échantillonnage, ce qui permet d'extraire une représentation composée de tokens discrets, chacun correspondant à des bandes de fréquence prédéfinies. Nous démontrons que cette représentation désentrelacée apporte une amélioration par rapport au modèle de base et ouvre de nouvelles possibilités pour l'interprétabilité du modèle.

\begin{figure}[t]
\centerline{\includegraphics[width=0.8\columnwidth]{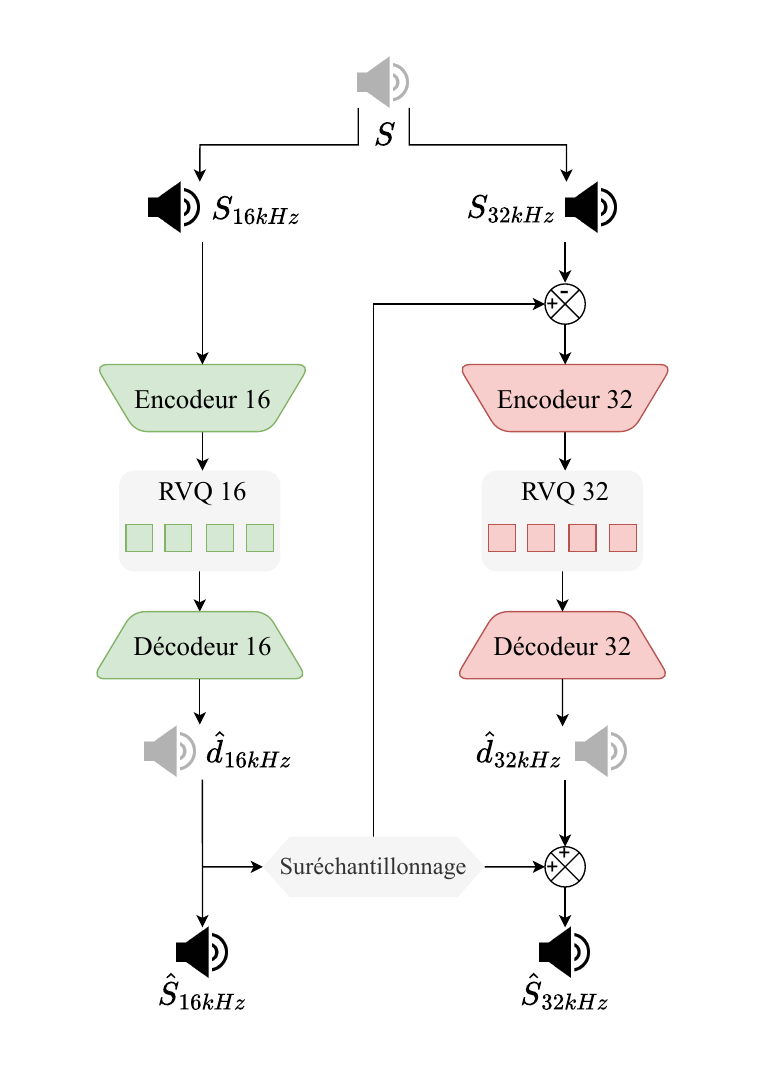}}
\caption{Le codec désentrelacé proposé.  
La branche $16~kHz$ reconstruit le signal $[0-8~kHz]$. La branche $32~kHz$ traite le résidu de $S_{32kHz}$ et $U(\hat{d}_{16kHz})$ pour produire le signal $[0-16~kHz]$ en additionnant les sorties de chaque branche.}
\label{fig-Um}
\end{figure}

\section{Codec désentrelacé}

Notre modèle présente une approche générique, facilement applicable, basée sur la décomposition fréquentielle des signaux audio, ce qui lui confère une grande interprétabilité. Notre approche consiste à adapter l'architecture d'un codec audio neuronal discret pour introduire une information fréquentielle dans la réprésentation extraite. 
Pour ce faire, nous divisons les spectrogrammes d'entrée en sous-bandes, qui sont ensuite traitées indépendamment (comme dans Luo \& al. \cite{luo_gull_2024}). Toutefois, cette méthode peut imposer des contraintes importantes, car chaque bande doit être reconstruite séparément et la combinaison des sous-bandes en un spectrogramme complet peut introduire des artefacts indésirables.

Notre approche n'y est pas sujette puisqu'elle s'appuie sur une décomposition douce des fréquences en travaillant dans le domaine temporel, et en exploitant le lien entre la fréquence d'échantillonnage d'un signal et son information spectrale. 
Nous ne considérerons qu'une version simple de notre approche avec deux branches (voir la figure \ref{fig-Um} pour une description schématique). Cette architecture simplifiée peut servir de preuve de concept et être facilement étendue à plusieurs branches fonctionnant chacune à des fréquences d'échantillonnage différentes. Les deux branches de notre codec audio neuronal se rapportent aux fréquences d'échantillonnages $16~kHz$ et $32~kHz$. Les signaux temporels d'entrée $(S_{16kHz},S_{32kHz})$ sont deux versions du même signal $S$, où $S_{16kHz}$ est échantillonné à $16~kHz$ (c'est-à-dire qu'il contient des informations dans la bande $[0-8~kHz]$) et $S_{32kHz}$ est échantillonné à $32~kHz$ (c'est-à-dire qu'il contient des informations dans la bande $[0-16~kHz]$).
Chaque branche produit un signal temporel, noté respectivement $\hat{d}_{16kHz}$ et $\hat{d}_{32kHz}$. Le signal reconstruit de $S_{16kHz}$, noté $\hat{S}_{16kHz}$, est directement obtenu à partir de la branche $16~kHz$, et~ : $\hat{S}_{16kHz} = \hat{d}_{16kHz}$. L'autre branche, qui fonctionne à $32~kHz$, prend en entrée le signal résiduel $S_{32kHz} - U(\hat{d}_{16kHz})$ où $U$ est l'opérateur de suréchantillonnage de $16~kHz$ à $32~kHz$. Le signal reconstruit de cette branche, noté $\hat{S}_{32kHz}$, est obtenu en additionnant directement les signaux issus des deux décodeurs : $\hat{S}_{32kHz} = U(\hat{d}_{16kHz}) + \hat{d}_{32kHz}$.

L'architecture en cascade proposée permet un désentrelacement doux : chaque branche est principalement conçue pour générer du contenu dans sa propre bande de fréquence mais aucune contrainte stricte n'impose cette séparation. Par conséquent, le décodeur de la branche $32~kHz$ peut également reconstruire du contenu à basse fréquence. Ce désentrelacement doux permet à la branche $32~kHz$ d'améliorer la reconstruction du signal en dessous de $8~kHz$ et d'atténuer les artefacts potentiels aux frontières entre les bandes de fréquences. D'autre part, un représentation issue d'une telle architecture peut démontrer un intérêt certain pour une tâche telle que l'extension de bande.

\section{Expériences}

\subsection{Données}

Pour nos expériences, nous avons utilisé les bases de données MUSDB18 \cite{musdb18} et Jamendo \cite{bogdanov2019mtg} qui rassemblent plus de 55 000 pistes de musique dans un jeu d'entraînement et 50 pistes musicales dans un jeu de test, initialement échantillonnées à $44,1~kHz$ avec une durée totale supérieure à 3 700 heures.

\subsection{Architecture}

Nous avons choisi de reproduire l'architecture du Descript Audio Codec (DAC \cite{kumar_high-fidelity_2023}), un état de l'art en matière de compression audio, dans une version dégradée (c'est-à-dire avec un RVQ ne contenant que quelques dictionnaires, ou en d'autres termes, avec un taux de compression élevé). Il se compose d'un codeur-décodeur convolutionnel, dans lequel est inséré un simple quantificateur vectoriel résiduel. 
Pour chacune des deux bandes de fréquences définies précédemment, nous avons reproduit une version de ce codec audio et adapté ses dimensions à la fréquence d'échantillonnage correspondante : nous avons choisi les rapports de compression des blocs encodeurs dans chaque branche de manière à ce que le rapport entre la fréquence d'échantillonnage de la branche et le taux de compression global soit constant d'une branche à l'autre.
Nous avons décidé de refléter la largeur de chaque bande de fréquence dans la profondeur du RVQ dans chaque branche. Cela a conduit à avoir quatre quantificateurs dans la branche $16~kHz$ et quatre autres quantificateurs dans la branche $32~kHz$. 

Pour la comparaison, nous avons également réentraîné le modèle DAC, uniquement sur MUSDB18 et Jamendo, à des fréquences d'échantillonnage de $16~kHz$ et $32~kHz$ en conservant les taux de compression et le débit binaire que nous avons définis dans notre modèle (il s'agit donc d'une version dégradée) : un débit binaire de $2~kbps$ pour le modèle de $16~kHz$ (taux de compression de 128) et un débit binaire de $4~kbps$ pour le modèle de $32~kHz$ (taux de compression de 128).

\begin{figure}[t]
\centerline{\includegraphics[width=0.9\columnwidth, trim={0 1cm 0 1.5cm},clip]{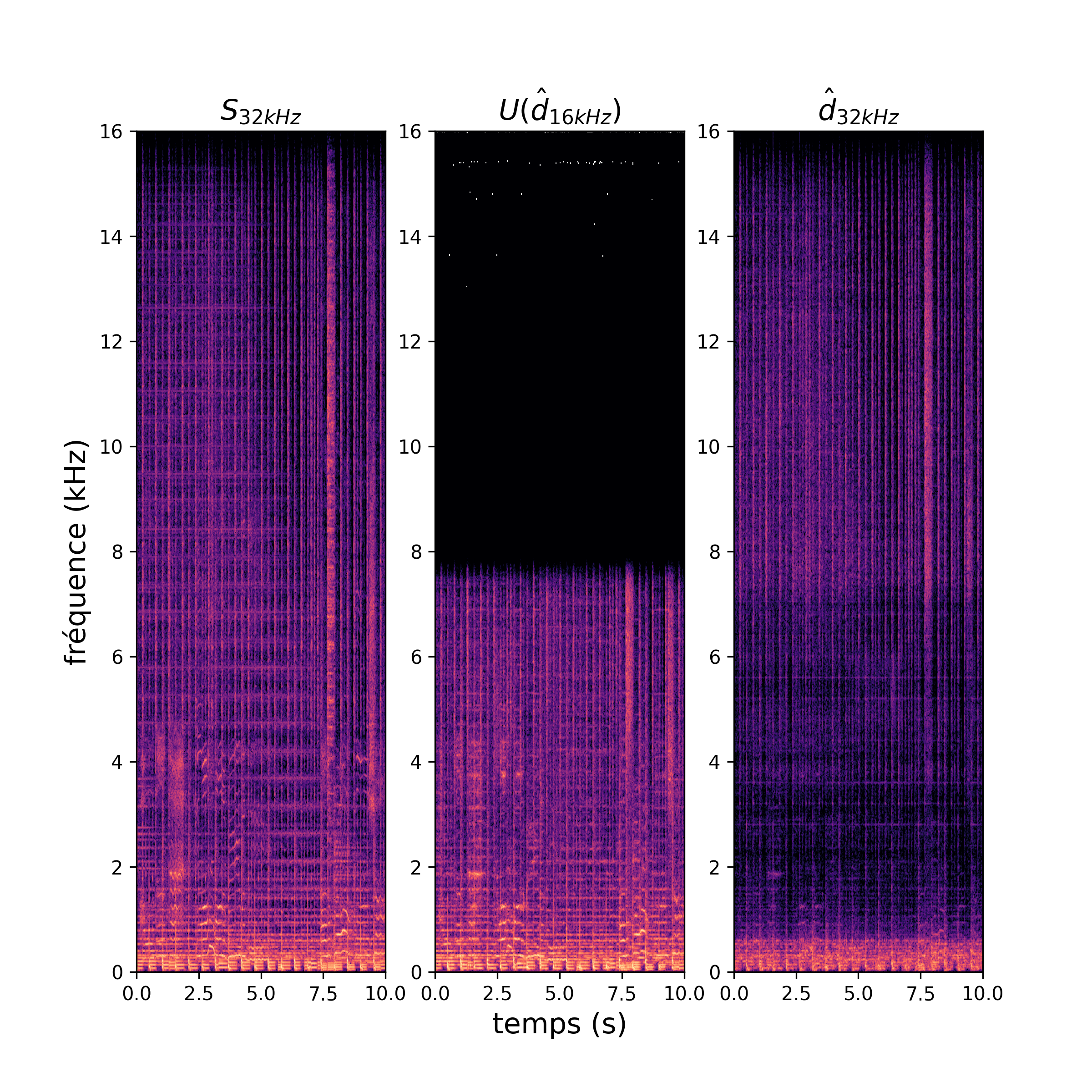}}
\captionsetup{justification=centering}
\caption{Spectrogrammes de $S_{32kHz}$, $U(\hat{d}_{16kHz})$ et $\hat{d}_{32kHz}$. \\
$U(\hat{d}_{16kHz})$ n'encode que l'information dans la bande $[0-8~kHz]$. $\hat{d}_{32kHz}$ a la plus grande partie de son énergie dans la bande $[8-16~kHz]$, même s'il porte également des informations résiduelles dans la bande inférieure.}
\label{fig-spec_dis}
\end{figure}

\subsection{Procédure d'entraînement}

Le codec ayant plusieurs branches, une légère adaptation de la procédure d'entraînement du modèle DAC \cite{kumar_high-fidelity_2023} a été nécessaire.
Toutes les branches sont d'abord entraînées en cascade, en commençant par la branche fonctionnant à la fréquence d'échantillonnage la plus basse : a) nous entraînons une branche, b) nous gelons ensuite ses paramètres et ceux de toutes les branches entraînées précédemment, et enfin c) nous entraînons la branche suivante. À la fin de cette formation en cascade, tous les poids de toutes les branches sont finement ajustés pour trouver un optimum global. Pour chaque branche associée à une fréquence d'échantillonnage $F$, nous lui adjoignons un réseau discriminant (en conservant la structure du discriminateur décrit dans \cite{kumar_high-fidelity_2023}), et nous calculons les coûts génératifs adverses ($\mathfrak{L}_{F}^{gen}$) et les coûts de \textit{matching features} adverses ($\mathfrak{L}_{F}^{fm}$), un coût de mel multi-échelles ($\mathfrak{L}_{F}^{mel}$) et les coûts des dictionnaires ($\mathfrak{L}_{F}^{cb}$) et un coût d'adhérence  ($\mathfrak{L}_{F}^{cmt}$) pour chaque branche. 

Pour les premières étapes d'apprentissage $16~kHz$ et $32~kHz$, nous définissons les coûts d'entraînement $\mathfrak{L}_{16kHz}^{total}$ et $\mathfrak{L}_{32kHz}^{total}$ :

\begin{equation}
\mathfrak{L}_{16kHz}^{total} = \sum_{\lambda}\left( \alpha_{16kHz}^{\lambda} \mathfrak{L}_{16kHz}^{\lambda}\right)
\label{eq-16_loss}
\end{equation}

\begin{equation}
\mathfrak{L}_{32kHz}^{total} = \sum_{\lambda} \left( \alpha_{32kHz}^{\lambda} \mathfrak{L}_{32kHz}^{\lambda}\right)
\label{eq-32_loss}
\end{equation}
Où $\lambda \in \{ gen, fm, mel, cb, cmt \}$. Nous avons choisi de définir le coût d'entraînement pour l'étape d'ajustement final ($\mathfrak{L}_{finetun}^{total}$) de la manière suivante :

\begin{multline}
\!\!\!\!\mathfrak{L}_{finetun}^{total} = \!\!\!\!\!\!\!\!\!\!\sum_{\lambda \in \{ gen, fm, mel \}} \!\! \frac{1}{2} \left( \alpha_{32kHz}^{\lambda} \mathfrak{L}_{32kHz}^{\lambda} + \alpha_{16kHz}^{\lambda} \mathfrak{L}_{16kHz}^{\lambda}\right) \\
+ \sum_{\lambda \in \{ cb, cmt \}} \left( \alpha_{32kHz}^{\lambda} \mathfrak{L}_{32kHz}^{\lambda} + \alpha_{16kHz}^{\lambda} \mathfrak{L}_{16kHz}^{\lambda}\right)
\label{eq-ft_loss}
\end{multline}

Nous calculons donc la moyenne des coûts génératifs, de feature matching et mel des deux branches, tandis que nous sommons simplement les coûts relatifs à la quantification, pour tenir compte de la différence de nature entre les coûts liés à la reconstruction (pour lesquelles les deux branches se chevauchent) et les coûts qui contraignent l'apprentissage des dictionnaires $16~kHz$ et $32~kHz$.
 
Nous avons choisi pour $F \in \{ 16kHz, 32kHz \}$, $\alpha_{F}^{gen} = 1. 0$, $\alpha_{F}^{fm} = 2.0$, $\alpha_{F}^{mel} = 15.0$, $\alpha_{F}^{cb} = 1.0$ et $\alpha_{F}^{cmt} = 0.25$, comme spécifié dans \cite{kumar_high-fidelity_2023}.

\section{Evaluation}

Nous avons également conservé les mêmes mesures d'évaluation que dans \cite{kumar_high-fidelity_2023} : coût de forme d'onde, coût TFTC, coût mel, rapport signal-distorsion échelle invariant (SI-SDR) tel qu'introduit dans \cite{le2019sdr}, et le score ViSQOL, une évaluation de la qualité audio perceptuelle introduite dans \cite{chinen2020visqol}.

L'informativité de la reconstruction a été mesurée grâce au rapport signal/distorsion (le SISDR \cite{le2019sdr} n'étant pas adapté au calcul par bande) exprimé en décibels, qui est défini pour $S_F$ un signal temporel échantillonné à $F$ et sa reconstruction $\hat{S}_F$ de la manière suivante :
$SDR(S_{F}, \hat{S}_{F}) =10 \log_{10}\left(\frac{||S_F||^2}{||S_F - \hat{S}_F||^2}\right)$.

Une évaluation de la qualité perceptive des reconstructions obtenues avec notre modèle désentrelacé par rapport à la référence DAC a été menée à travers un test MUSHRA \cite{schoeffler_webmushra_2018}. Treize participants ont du noter de 0 (mauvais) à 100 (excellent) la qualité perceptive de quatre versions d'un même extrait audio : une référence, une version encodée par DAC, une version encodée par notre modèle et une ancre (filtre passe bas à $3.5~kHz$). Ce test a été effectué pour les reconstructions à $16~kHz$ et $32~kHz$. Deux ensembles de 6 extraits ont été tirés au hasard de l'ensemble de test, pour les deux fréquences d'échantillonnage.

\section{Résultats}

\begin{table}[t]
\caption{\centering Métriques de reconstruction du codec désentrelacé}
\begin{center}
\begin{tabular}{|c||c|c||c|c|}
\hline
\'Echantillonage & \multicolumn{2}{|c||}{$16 000~Hz$} & \multicolumn{2}{|c|}{$32 000~Hz$} \\
\hline
Débit & \multicolumn{2}{|c||}{$2~kbps$} & \multicolumn{2}{|c|}{$4~kbps$} \\
\hline
Modèle & DAC & CD & DAC & CD \\
 & & (proposé) &  & (proposé) \\
\hline
mel ($\downarrow$)&1.08&\textbf{0.95}&0.90&\textbf{0.80}\\
stft ($\downarrow$)&2.67&\textbf{2.52}&2.28&\textbf{2.14}\\
reconstruction ($\downarrow$)&0.072&\textbf{0.066}&0.060&\textbf{0.05}\\
SI-SDR ($\uparrow$)&2.97&\textbf{3.90}&5.00&\textbf{6.05}\\
ViSQOL ($\uparrow$)&4.08&\textbf{4.25}&3.97&\textbf{4.22}\\
\hline
\end{tabular}
\label{tab-reco_met}
\end{center}
\end{table}

\begin{table}[t]
\caption{ \centering Notes moyennes de qualité perceptive obtenues par le test MUSHRA ($\pm$ écart type)}
\begin{center}
\begin{tabular}{|c||c||c|c||c|}
\hline
& Référence & DAC & CD & Ancre \\
&  &  & (proposé) &  \\
\hline
$16~kHz$&95 $\pm$ 9&31 $\pm$ 18&\textbf{53} $\pm$ 20&37 $\pm$ 25\\
$32~kHz$&96 $\pm$ 5&49 $\pm$ 20&\textbf{66} $\pm$ 19&38 $\pm$ 24\\
\hline
\end{tabular}
\label{tab-mushra}
\end{center}
\end{table}

\begin{table}[t]
\caption{Désentrelacement - SDR par bande de fréquences}
\begin{center}
\begin{tabular}{|c|c||c|c|}
\hline
\multicolumn{2}{|c||}{\rule{0pt}{2.6ex} $\hat{d}_{32kHz}$ vs $S_{32kHz}$ \rule[-1.2ex]{0pt}{0pt}} & \multicolumn{2}{|c|}{\rule{0pt}{2.6ex} $\hat{S}_{32kHz}$ vs $S_{32kHz}$ \rule[-1.2ex]{0pt}{0pt}} \\
\hline
$[8-16~kHz]$ & 5.85 & $[7.9-8.1~kHz]$ & 5.61 \\
$[0-8~kHz]$ & 3.80 & $[0-16~kHz] $& 5.67\\
\hline
\end{tabular}
\label{tab-SDR}
\end{center}
\end{table}

Les résultats affichés dans la table \ref{tab-reco_met} montrent une légère amélioration des performances de notre modèle à $16~kHz$ par rapport à la référence, liée à l'étape d'ajustement fin. De même, les mesures de reconstruction observées pour la reconstruction globale à $32~kHz$ indiquent une amélioration de la qualité de la reconstruction par rapport à la version $32~kHz$ du modèle DAC.

Les mesures perceptives obtenues par le test MUSHRA et résumées dans la table \ref{tab-mushra} confirment l'observation que nous avons faite avec les métriques de reconstruction : pour les reconstructions $16~kHz$ et $32~kHz$, notre modèle est plus performant que la référence, même si les écarts types élevés soulignent que les écarts ne sont pas significatifs.  

Le désentrelacement de l'information portée par les tokens extraits de notre codec a été étudié en comparant la sortie de chaque branche ($U(\hat{d}_{16kHz})$ et $\hat{d}_{32kHz}$) au signal d'entrée $S_{32kHz}$. Les spectrogrammes de ces signaux sont représentés sur la figure \ref{fig-spec_dis}.
La reconstruction issue de la branche $16~kHz$ n'encode des informations que dans les basses fréquences, car la fréquence d'échantillonnage ne permet pas de reconstruire au-delà de $8~kHz$. Quant à la reconstruction de la branche $32~kHz$, la plupart des informations sont portées dans les hautes fréquences, même si le désentrelacement doux apporte des corrections dans la bande des basses fréquences. La table \ref{tab-SDR} rassemble des valeurs de SDR entre $S_{32kHz}$ et $\hat{d}_{32kHz}$, en divisant le calcul entre la bande $[0-8~kHz]$ et la bande $[8-16~kHz]$.
L'information du signal reconstruite par la branche $32~kHz$ est, en effet, beaucoup plus corrélée à $S_{32kHz}$ dans les hautes fréquences (avec un SDR de $5.85~db$), même si certaines informations pertinentes sont également encodées dans les basses fréquences (avec un SDR plus faible de $3.80~db$), grâce au désentrelacement doux. 

Les résultats de la deuxième partie da la table \ref{tab-SDR} soulignent également que le SDR calculé dans une bande étroite autour de la fréquence de coupure commune des deux sous-bandes ($8~kHz$) est presque identique au SDR moyen sur l'ensemble du spectrogramme, indiquant l'absence d'artefacts dans cette zone.

\section{Conclusion}

Dans ce travail, nous introduisons une approche pour concevoir un codec audio neuronal en incorporant une décomposition fréquentielle des signaux d'entrée, qui facilite le désentrelacement des représentations discrètes. Nous démontrons qu'avec une amélioration de la reconstruction, cette représentation favorise également l'interprétabilité des représentations extraites.


\bibliography{abrv, biblio}


\end{document}